\documentstyle[12pt]{article}
\def\cqkern#1#2#3{\copy255 \kern-#1\wd255 \vrule height #2\ht255 depth
   #3\ht255 \kern#1\wd255}

\def\a{\begin{eqnarray}}
\def\b{\end{eqnarray}}

\begin{document}
\centerline{\LARGE On the Linearized Artin Braid Representation}
\vspace{1truecm}
\vskip0.5cm
\centerline{\large F. Constantinescu}
\vskip.5cm
\centerline{Fachbereich Mathematik}
\centerline{Johann Wolfgang Goethe Universit\"{a}t Frankfurt,}
\centerline{Robert Mayer Str. 10,
D-6000 Frankfurt a. M. 1}
\vskip.5cm
\centerline{\large F. Toppan}
\vskip.5cm
\centerline{Physikalisches Institut}
\centerline{Universit\"{a}t Bonn}
\centerline{Nussallee 12, D-5300 Bonn 1}
\vskip1.5cm
\centerline{\bf Abstract}
\vskip.5cm
We linearize the Artin representation of the braid group given
by (right) automorphisms of a free group providing a linear
faithful representation of the braid group. This result is
generalized to obtain linear representations for the coloured braid
groupoid and pure braid group too. Applications to some areas of
two-dimensional physics are discussed.

\vfill\eject

\section{Introduction}

Let us consider a free group $F_n$ of rank $n$ with generators
$x_1, x_2,...,x_n$.
Artin \cite{artin1,artin2} proved that the braid group
$B_n$ with generators
$\sigma_1,...,\sigma_{n-1}$ and defining relations
\begin{eqnarray}
{\sigma_i}{\sigma_{i+1}}{\sigma_i}&=&{\sigma_{i+1}}{\sigma_i}{\sigma_{i+1}}
\nonumber \\
{\sigma_i}{\sigma_j}&=&{\sigma_j}{\sigma_i}\quad for \quad |i-j|\geq 2
\label{relations}
\end{eqnarray}
has a faithful representation
as a group of right automorphisms $ {\hat\sigma}_i$,\\
$i=1,...,n-1$
 of $F_n$ given by
\a
&&{\hat\sigma}_i:
\left\{
\begin{array}{lll} {x}_i&\mapsto&{x}_i\cdot
{x}_{i+1}\cdot {{x}_i}^{-1} \\
{x}_{i+1}&\mapsto& {x}_i  \\
{x}_j&\mapsto& {x}_j \quad for \quad j\neq i, i+1
 \end{array}
\right.
\label{artin1}
\b
On the other hand
Magnus \cite{magnus} obtained a class of representations of free groups which
can be used to get
in some cases representations of subgroups of the automorphism group of
$F_n$. In this paper we prove that the linearization of the Artin
transformations (\ref{artin1}) using the concept of
the Magnus representation leads
to a faithful
representation of the braid group which was already introduced
in some different context in
\cite{ConLue}:\\
Let ${\cal J}B_{n}$ be the ring over the braid
group $B_{n}$ with integer coefficients.
The representatives of the braid group generators are the matrices
\a
({\hat\sigma}_i)_{jk} &=&{\bf 1}_{(i-1,i-1)}\cdot \sigma_i \oplus
\left(
\begin{array}{cc}
\alpha_i & \beta_i \\
\sigma_i & 0
\end{array}
\right) \oplus {\bf 1}_{n-2-i,n-2-i}\cdot \sigma_i
\label{represent}
\b
with
\a
\alpha_i &=& \sigma_i (  1-\theta_i\theta_{i+1}{\theta_i}^{-1})
\nonumber\\
\beta_i &=& \sigma_i \theta_i
\label{zzz}
\b
and
\a
\theta_i &=& {\sigma_1}^{-1}...{\sigma_{i-1}}^{-1} {\sigma_i}^2{\sigma_{i-1}}
...\sigma_1
\label{definit}
\b
The fact that (\ref{represent}) provides a representation
of $B_n$ with values in the braid ring ${\cal J}B_{n}$ can be seen from
the following equalities in the braid ring:
\a
\alpha_i\sigma_{i+2}\alpha_i +\beta_i\alpha_{i+1}\sigma_{i+1} &=&
\sigma_{i+2}\alpha_i\sigma_{i+2}\nonumber\\
\alpha_i\sigma_{i+2}\beta_i &=& \sigma_{i+2}\beta_i\alpha_{i+1}\nonumber \\
\beta_i\beta_{i+1}\sigma_{i+1} &=& \sigma_{i+2}\beta_i\beta_{i+1}
\nonumber \\
\sigma_{i+1}\sigma_{i+2}\alpha_i &=&
\alpha_{i+1}\sigma_{i+1}\sigma_{i+2}\nonumber\\
\sigma_{i+1}\sigma_{i+2}\beta_i &=& \beta_{i+1}\sigma_{i+1}\sigma_{i+2}
\label{identit}
\b
A full tower of linear representations of the braid group is produced
by iteratively replacing the old generators with the new representatives.
The trivial diagonal representation $ \sigma_i = q\in {\bf R}$ generates
for instance the Burau
representation which again, introduced in (\ref{represent}), produces
a new representation and so on. The representation (\ref{represent})
is reducible as a consequence of the fact that the element
$x_1\cdot x_2\cdot...\cdot x_n\in F_n$ is left
invariant under Artin transformations.\\
Using the braid relations $\alpha_i$ can be reexpressed to
$\alpha_i = (1-\theta_i)\sigma_i$ \cite{prilaw}.\\
Another proof of the fact that (\ref{represent}) is a representation
of the braid group
using homology was indicated to the authors by
R. Lawrence \cite{prilaw};
it is based on her very interesting thesis \cite{Law2}.
It seems that ideas similar to ours can also be found in the book \cite{Burde}
and go back to Artin.\\
Here we also extend the construction which leads to (\ref{represent})
to the case of the coloured braid
groupoid and pure braid group.\\
The representation (\ref{represent})
and especially its iterations have applications which are listed here:
determination of braid and monodromy properties of generalized
hypergeometric integrals,
characterization of braid and monodromy representations
through bilinear forms
\cite{ConLue2,flu},
contour methods
in two dimensional conformal quantum field theory (especially perturbation
theory where the vertex structure is partially spoiled \cite{flu}) and
quantum groups \cite{LueTop,Lue}.
These applications will be briefly discussed in section 5 of this paper. For
more details see \cite{ConLue,LueTop,Lue}.\\
Concerning the organization of the paper,
we put in section 2 some remarks about
the free differential calculus and the Magnus representation of free groups
and braid group.
In section 3 we prove that (\ref{represent}) provides a linear faithful braid
valued representation of the braid group; in section 4 we extend the results
to the coloured and pure braid group cases.

\section{The Magnus representation}

We follow \cite{birman} with some minor modifications.\\
Let $\Phi$
be an arbitrary homomorphism action on
the free group $F_n$ and let ${F_n}^{\Phi}$ denote
the image of $F_n$ under $\Phi$. Let
${{\cal J}F_n}^{\Phi}$ denote the group ring of
${F_n}^{\Phi}$ with integer coefficients. Elements in ${{\cal J}F_n}^{\Phi}$
are formal linear combinations of the form
$ \sum {a_g}g$ ($g\in{ F_n}^{\Phi}$, $a_g\in {\cal J}$).\\
The operations in ${{\cal J}F_n}^{\Phi}$ are defined as follows:
\a
\sum {a_g}g +\sum {b_g}g &=& \sum {(a_g+b_g)}g\nonumber\\
(\sum {a_g}g)\cdot (\sum {b_g}g) &=& \sum_g(\sum_h{a_{gh^{-1}}}b_h)g
\b
There is a well defined mapping ${\partial\over\partial x_j}$,
$j=1,...,n$, called
the free differential
\begin{eqnarray}
{\partial\over\partial x_j} &:& {\cal J}F_n \rightarrow {\cal J}F_n
\end{eqnarray}
given by
\a
{\partial\over\partial x_j}({x_{\mu_1}}^{\epsilon_1}...
{x_{\mu_r}}^{\epsilon_r})&=& \sum_{i=1}^r \epsilon_i\delta_{\mu_i,j}
{x_{\mu_1}}^{\epsilon_1}...{x_{\mu_i}}^{{\textstyle{1\over 2}}(\epsilon_i-1)}
\nonumber\\
{\partial\over\partial x_j}(\sum a_g g) &=& \sum a_g{\partial
g\over\partial x_j}
\b
where $g\in F_n$, $a_g\in{\cal J}$, $\epsilon_i =\pm1$.
\\
We have the rules
\a
&& \begin{array}{ll} i) & {\partial x_i\over\partial x_j}
=\delta_{i,j}\\
 ii) & {\partial{x_i}^{-1}\over\partial x_j} = -\delta_{i,j} {x_i}^{-1}
\\ iii) & {\partial (wv)\over\partial x_j}
= {\partial w\over\partial x_j} v^t +
w{\partial v\over\partial x_j}\\
 iv) & chain\quad rule\end{array}
\b
In $iii)$ $v^t$ means the sum of the integer-coefficients in $v$ and $iv)$
can be formulated as follows:
let $v_1,...,v_n$ be another system of generators in $F_n$. They are words
$v_i(x_1,...,x_n)$, $i=1,...,n$ in $F_n$. We have
\a
{\partial\over\partial x_j}w (v_1,...,v_n) &=& \sum_{ k=1}^n
{\partial w\over\partial v_k}{\partial v_k\over\partial x_j}
\nonumber
\b
Remark the unusual form of $ii)$ and $iii)$.\\
In section 4 we will consider the monodromy subgroup
of the pure braid group, given by the generators
$\theta_i$ of formula (\ref{definit}).
Every free group is isomorphic to the group
generated by $\theta_i $. Via this identification we
are able to
consider the multiplication operation of elements of a free group by elements
of the braid group $B_n$.\\
We are going now to recall to the reader the Magnus representations of a
free group and of the braid group.
Let $S_n$ be a free abelian semigroup with basis $s_1,...,s_n$ and let
$A ({\cal R},S_n)$ be the semigroup ring of $S_n$ with respect to the ring
${\cal R}$.
Let $\Phi$ be a homomorphism acting on the free group $F_n$; for $w\in F_n$
let
$w^{\Phi}$ be the image of $w$ under $\Phi$. Then the mapping
$w\rightarrow (w)^{\Phi}$,
\a
(w)^{\Phi} &=& \left(
\begin{array}{cc}
w^{\Phi} &\sum_{j=1}^n ({\partial w\over\partial x_j})^{\Phi} s_j\\
0& 1
\end{array}
\right)
\b
with the entries of $ (w)^{\Phi}$
in $A({\cal J}{F_n}^{\Phi},S_n)$,
is a representation of $F_n$ called the Magnus $\Phi$-representation.
It is faithful if $F_n$ is abelian; otherwise its kernel
is the commutator subgroup of $ker\Phi$ \cite{birman}.
Now let $\cal{A}$ be any group of right (automorphisms) of $F_n$ such that
\a
x\Phi & = & x\alpha\Phi
\label{condition}
\b
for each $ x\in F_n$,$\alpha\in{\cal A}$.
Then the matrix
\a
||\alpha||^{\Phi} &=&\left(
{\partial (x_i\alpha )\over \partial x_j} \right)^{\Phi}
\b
with entries in ${{\cal J}F_n}^{\Phi}$
defines a linear representation of ${\cal A}$ called again
Magnus representation. If the automorphism group ${\cal A}$ is
the Artin automorphism (\ref{artin1}), then we get the Magnus representation
of the braid
group. If $\Phi$ is defined by
$ x_i\Phi = q\in {\bf R} $, $1\leq i\leq n$, we recover the
Burau representation as
a special case of the Magnus representation.
The proof of the above assertions is based on the chain rule of the
free differential
calculus and uses heavily the condition (\ref{condition}).
In fact it turns out that this condition is rather
restrictive because in the braid
case it implies that $x_i\Phi$ must be independent of $i$.\\
In the next section we will drop out the condition (\ref{condition})
in the particular case when ${\cal A}$ coincides with the Artin's
automorphism group
and prove what we call the braid valued generalization of the Magnus
representation for the braid group.

\section{Braid valued representation of the braid group}

Let ${\tilde{F_n}}$ be now a copy of $F_n$ wih generators $s_i$.
Certainly there will be a (generalized)
Magnus representation of $F_n$ given by
$w\rightarrow (w)$, $w\in F_n$ and
\a
(w) &=& \left(
\begin{array}{cc} w & \sum_{j=1}^n {\partial w\over
\partial x_j} s_j\\
0 & 1 \end{array}\right)
\b
where the entries of $(w)$ are in the free ring of ${\tilde F}_n$
with respect to the ring ${\cal J}F_n$.
For the generators this representation reduces to
\a
(x_i) &=& \left(
\begin{array}{cc}
x_i & s_i\\
0 & 1 \end{array}\right)
\b
At the first glance this representation seems not to be very interesting,
however we will prove that it provides a representation of the braid
group via
Artin automorphism.\\
We identify the free group $F_n$
with the monodromy subgroup of
the pure braid group via the identification
of the generators $x_i$ and $\theta_i$.
For the monodromy group the Artin transformations are equivalent to the
adjoint action of the generators $\sigma_i$:
\a
{\sigma_i}^{-1}\theta_k\sigma_i &=& {\hat\sigma}_i(\theta_k)\quad for\quad any
\quad 1\leq i,k\leq n
\label{adjoint}
\b
where
\a
{\hat\sigma}_i ({\theta}_i)&=&{\theta}_i\cdot
{\theta}_{i+1}\cdot {{\theta}_i}^{-1} \nonumber \\
{\hat\sigma}_i ({\theta}_{i+1})&=& {\theta}_i \nonumber \\
{\hat\sigma}_i ({\theta}_j)&=& {\theta}_j \quad for \quad j\neq i, i+1
\nonumber
\b
For given $i$ the right hand side of (\ref{adjoint}) provides
a new system of generators of the monodroy group.\\
We go now from the free group $F_n$ to the free group $F_n'$ whose generators
$\theta_j'$ are the images under the Artin transformation ${\hat\sigma}_i$
of the $F_n$ generators:
\a
\theta_j'&=& {\sigma_i}^{-1} \theta_j\sigma_i\quad ,\quad 1\leq i,j\leq n
\nonumber
\b
Let us introduce for a given $i$ the jacobian map $J$ of the free
differential calculus
\a
J &:& {\cal J}F_n\rightarrow {\cal J}F_n
\nonumber\\
J&=& \left( {\partial {\hat \sigma}_i(\theta_j)\over\partial\theta_k}\right)
\label{jaco}
\b
which is the linearization of the Artin transformation.
We proceed in the same way as before going from the free group $F_n'$
to $F_n''$ with generators
\a
\theta_j''&=& {\sigma_l}^{-1}\theta_j'\sigma_l\quad,\quad 1\leq l,j\leq n
\nonumber
\b
The linearization of the Artin relations (applied to $\theta_j'$) is given
by the jacobian $J': {\cal J}F_n'\rightarrow {\cal J} F_n'$
defined as (\ref{jaco}) with the replacement $\theta_j\rightarrow\theta_j'$.\\
It is possible to transport the linear mapping $J'$ to a linear mapping in
${\cal J}F_n$ by using the commutativity of the following
diagram
\a
&& \begin{array}{ccc}
{\cal J}F_n& \begin{array}{c} J\\
\longrightarrow \end{array} & {\cal J}F_n \\
\downarrow &\quad &\downarrow\\
{\cal J}F_n' &\begin{array}{c} J'\\
\longrightarrow \end{array} &{\cal J}F_n'
\end{array}
\label{diagram}
\b
where the vertical arrow denotes the adjoint map
$ad_{\sigma}$:
\a
ad_{\sigma} : \beta \mapsto {\sigma_i}^{-1} \beta\sigma_i
\in {\cal J}F_n'\quad,\quad\beta\in{\cal J}F_n
\b
The above diagram makes possible to write
\a
J'&=& {\sigma_i}^{-1}J\sigma_i
\b
We get, using the braid relations:
\a
({\sigma_j}^{-1}J_i\sigma_j)J_j &=& ({\sigma_i}^{-1}J_j\sigma_i)J_i
\quad ,\quad |i-j|\geq 2
\label{up}
\b
and
\a
&&\relax [{\sigma_{i+1}}^{-1}({\sigma_i}^{-1}J_{i+1}\sigma_i)\sigma_{i+1}]
({\sigma_{i+1}}^{-1}J_i\sigma_{i+1})J_{i+1}=\nonumber\\
&&\relax [{\sigma_i}^{-1}({\sigma_{i+1}}^{-1}J_i\sigma_{i+1})\sigma_i]
({\sigma_i}^{-1}
J_{i+1} \sigma_i) J_i
\label{down}
\b
(\ref{up}) and (\ref{down}) give, by inserting factors ${\sigma}^{-1}\sigma$,
\a
{\sigma_j}^{-1}{\sigma_{i}}^{-1}\sigma_iJ_i\sigma_{j}J_j &=&
{\sigma_i}^{-1}{\sigma_j}^{-1}{\sigma_j}
J_{j} \sigma_i J_i\quad ,\quad |i-j|\geq 2
\label{inter1}
\b
and
\a
&&{\sigma_{i+1}}{\sigma_i}{\sigma_{i+1}}{\sigma_{i+1}}^{-1}{\sigma_i}^{-1}
J_{i+1} {\sigma_i}{\sigma_{i+1}}{\sigma_{i+1}}^{-1}J_i{\sigma_{i+1}}
J_{i+1} =\nonumber\\
&&{\sigma_{i}}{\sigma_{i+1}}{\sigma_{i}}{\sigma_{i}}^{-1}{\sigma_{i+1}}^{-1}
J_{i} {\sigma_{i+1}}{\sigma_{i}}{\sigma_{i}}^{-1}J_{i+1}{\sigma_{i}}
J_{i}
\label{inter2}
\b
The relations (\ref{inter1},\ref{inter2}) can be written now by using
again the braid relations as follows
\a
(\sigma_iJ_i)(\sigma_jJ_j) &=& (\sigma_jJ_j)(\sigma_iJ_i) \quad ,\quad
|i-j|\geq 2 \\
(\sigma_iJ_i)(\sigma_{i+1}J_{i+1})(\sigma_iJ_i) &=&
(\sigma_{i+1} J_{i+1})(\sigma_iJ_i)(\sigma_{i+1}J_{i+1})
\b
This shows that $\sigma_i\rightarrow \sigma_iJ_i$ is a linear representation
of the braid group. By computing the matrix elements
$ ({\partial{\hat\sigma}_i(\theta_j)\over\partial\theta_k})$ of $J_i$ following
the rule
of the free differential calculus, we get (\ref{represent}).
As an example take
\a
{\partial{\hat\sigma}_i(\theta_i)\over\partial\theta_i}&=&
{\partial\over\partial\theta_i}(\theta_i\theta_{i+1}{\theta_i}^{-1})=
{\partial\theta_i\over\partial\theta_i}+\theta_i{\partial\over\partial\theta_i}
(\theta_{i+1}
{\theta_i}^{-1}) =\nonumber\\
&=& 1+\theta_i\theta_{i+1} {\partial\over\partial\theta_i}({\theta_i}^{-1})=
1-\theta_i\theta_{i+1}\theta_i
\b
Since the representation of the braid group through the Artin
right authomorphism is
faithful and
$ {\sigma_i}^{-1} \theta_j\sigma_i$ for given $i$
is an isomorphism of $F_n$ on $F_n'$,
it follows that the braid valued
representation (\ref{represent}) is a linear faithful representation.

\section{The coloured case and the pure braid group}

In this section we apply the previous construction
to the coloured case (representations of the coloured
braid grupoid) and to the pure braid group.\\
We introduce first the coloured generators $\sigma_i (\lambda ,\mu )$
(where $\lambda ,\mu,...$ can be looked as colours attached to strings)
of the
coloured braid groupoid ${B_n}^c$; they satisfy the relations
\a
\sigma_i(\lambda ,\mu ){\sigma_{i+1}} (\lambda ,\nu )\sigma_i(\mu ,\nu )
&=&
\sigma_{i+1} (\mu ,\nu ) \sigma_i (\lambda ,\nu )\sigma_{i+1} (\lambda ,\mu )
\b
and
\a
\sigma_i (\lambda ,\mu ) \sigma_j (\nu ,\rho ) &=&
\sigma_j (\nu ,\rho ) \sigma_i(\lambda ,\mu ) , \quad for \quad |i-j|\geq 2
\b
The inverses are introduced through the relation
\a
\sigma_i (\lambda ,\mu ){\sigma_i}^{-1} (\mu ,\lambda ) &=& 1
\b
The monodromies $\theta_i$ are straightforwardly generalized to
the coloured case as
$\theta_i^\rho (...,\lambda ,\mu ) $, where $\lambda$ , $\mu $ are the
last enclosed colours and $\rho$ is the colour associated to $\theta_i$:
\a
\theta_i^{\rho} (...,\lambda ,\mu ) &=& ...
{\sigma_{i-2}}^{-1} (\rho ,\lambda )
\sigma_{i-1}(\rho ,\mu )
\sigma_{i-1} (\mu ,\rho )\sigma_{i-2}(\lambda ,\rho )...
\b
The considerations of section 3 can be generalized to the coloured
braid grupoid and
provide a matrix representation of the generators generalizing
(\ref{represent}).
The matrix
${B_i^c} (\lambda ,\mu )$
is obtained from (\ref{represent})
by replacing $\sigma_i\rightarrow\sigma_i (\lambda ,\mu )$,
$\theta_i \rightarrow\theta_i^\rho (...,\mu ,\lambda )$:
\a
&& (B_i^c)_{jk}{(...,\lambda ,\mu )}\equiv
(B_i^c)_{jk}{(\lambda ,\mu )} = \nonumber\\
&& {\bf 1}_{(i-1,i-1)}\cdot
\sigma_i(\lambda ,\mu )
\oplus
\left(
\begin{array}{cc}
\alpha_i (\lambda ,\mu ) & \beta_i (\lambda ,\mu ) \\
\sigma_i{\lambda ,\mu} & 0
\end{array}
\right) \oplus {\bf 1}_{n-2-i,n-2-i}\cdot\sigma_i(\lambda ,\mu )
\nonumber\\
\label{represent2}
\b
where $\alpha_i (\lambda ,\mu )$, $\beta_i (\lambda ,\mu )$
can be easily read from (\ref{zzz}).\\
In the particular case in which the generators
$\sigma_i (\lambda ,\mu )$ are represented by $q_i\in {\bf R}$,
we get from the matrices ${B_i^c} (\lambda ,\mu )$
a representation of the coloured braid grupoid;
when specialized to the generators of the pure braid group this representation
coincides with the one discovered by Gassner \cite{birman}.
It follows that the Gassner matrices satisfy not only the pure braid relations
but also
\a
&&
\left( \begin{array}{ccc}
1&0&0\\
0&1-t_{i+1}&t_i\\
0&1&0
\end{array}  \right)
\left(
\begin{array}{ccc}
1-t_{i+2}&t_i&0\\
1&0&0\\
0&0&1
\end{array}\right)
\left(
\begin{array}{ccc}
1&0&0\\
0&1-t_{i+2}&t_{i+1}\\
0&1&0
\end{array}
\right) =
\nonumber\\
&&
\left(
\begin{array}{ccc}
1-t_{i+2}&t_{i+1}&0\\
1&0&0\\
0&0&1
\end{array}
\right)
\left(
\begin{array}{ccc}
1&0&0\\
0&1-t_{i+2}&t_i\\
0&1&0
\end{array}
\right)
\left(
\begin{array}{ccc}
1-t_{i+1}&t_i&0\\
1&0&0\\
0&0&1
\end{array}
\right)
\label{yangb}
\b
and iterated relations of this kind, (see also \cite{kauff}).\\
Equation  (\ref{yangb}) remembers the Yang-Baxter relation,
but for the fact that
it lives on a direct sum instead of a tensor product.
The problem of promoting a braid group representation from acting on a
direct sum to a braid group representation on a tensor product space is of
particular interest in the theory of quantum groups. A particular nice
example was provided by L. Kauffman and H. Saleur \cite{kauff} in
connection with the quantum supergroup $U_qgl(1,1)$. The tensor product space
is constructed in this case with the help of the exterior algebra over the
direct sum space.\\
We come now to the pure braid group $P_n$. It can be algebraically
defined as the subgroup of $B_n$ having generators
\a
{\theta_k}^{(i)}&=& {\sigma_i}^{-1}...{\sigma_{k-2}}^{-1}{\sigma_{k-1}}^2
\sigma_{k-2}...\sigma_i
\b
for $1\leq i<k\leq n$
and defined relations which appear for instance in \cite{birman}, p. 29.
Here we will consider the pure braid group $P_{n+1}$ with the extra
generators
\a
{\theta_k}^{(0)}&=& {\sigma_0}^{-1}...{\sigma_{k-2}}^{-1}{\sigma_{k-1}}^2
\sigma_{k-2}...\sigma_0\quad,\quad 1< k\leq n
\nonumber
\b
where $\sigma_0$ is the extra generator of $B_{n+1}$ as compared to $B_n$.
The same procedure as before allows us to construct a representation
of $P_n$ realized with matrices having entries
in ${\cal J}P_{n+1}$.
It is convenient to express the $n\times n$ representative matrices
${\hat\theta_k}^{(i)}$ of the
generators in terms of their action on the $n$ generators $y_j$ of a
(right-) module over the ring ${\cal J}P_{n+1}$. It turns out
that:\\
i) For $j<i$ or $j>k$ we get
\a
{{\hat\theta}_k}^{(i)}&:& y_j \mapsto{{\theta}_k}^{(i)}y_j\nonumber
\b
this relation being trivial.\\
ii)
\a
{{\hat\theta}_k}^{(i)}&:& y_k\mapsto {{\theta}_k}^{(i)}(1-{{ \theta}_i}^{(0)}
{{\theta}_k}^{(0)}{{\overline \theta}_i}^{(0)})y_i+\nonumber\\
&&
{{\theta}_k}^{(i)}{{\theta}_i}^{(0)}y_k\nonumber
\b
iii) for $ i\leq j<k $
\a
{{\hat\theta}_k}^{(i)}&:& y_j\mapsto
A_{(i,j,k)}y_j+B_{(i,j,k)}y_k+C_{(i,j,k)}y_i\nonumber
\b
where
\a
A_{(i,j,k)}
&=&{{\overline \theta}_i}^{(0)}{{\theta}_k}^{(i)}
{{\theta}_i}^{(0)}\nonumber
\b
\a
B_{(i,j,k)}
&=&[ 1-{{\overline \theta}_i}^{(0)}
-
{{\theta}_j}^{(0)}
+{{\theta}_j}^{(0)}
{{\overline \theta}_i}^{(0)}
]\cdot\nonumber \\
&& \cdot
{{\theta}_k}^{(i)}{{\theta}_i}^{(0)}\nonumber
\b
\a
C_{(i,j,k)}
&=&
[1-
{{\theta}_j}^{(0)}
+{{\theta}_j}^{(0)}{{\overline \theta}_i}^{(0)}
]{ \theta_k}^{(i)}+\nonumber \\
&& {{\theta}_j}^{(0)}{{\theta}_k}^{(i)}{{\theta}_i}^{(0)}-
{{\theta}_k}^{(i)}
{{\theta}_i}^{(0)}{{\theta}_k}^{(0)}{{\overline \theta}_i}^{(0)}-\nonumber \\
&&-{{\overline \theta}_i}^{(0)}
{{\theta}_k}^{(i)}{{\theta}_i}^{(0)}{{\theta}_j}^{(0)}
{{\overline \theta}_i}^{(0)}\nonumber
\b
In ii) and iii) the inverse of ${\theta_i}^{(k)}$ has been denoted by
${{\overline\theta}_i}^{(k)}$ for tipografical reasons.

\section{Applications}

A geometric visualization of the representation
(\ref{represent}) was given in \cite{ConLue}. We start by describing
the connection of (\ref{represent}) with
the realization of the braid group on analytic functions
with isolated branch points singularities (for details see \cite{ConLue}).
Let
\a
M_n^> &=& \{(z_1,...,z_n) : |z_i|>|z_j|,\quad if\quad i<j,\quad
-i\pi<arg z_k\leq i\pi\}
\b
be a simply connected subset of
${\bf C}^n$.
Let $\{f_j, j\in J \}$ be a family of holomorphic functions
in $M_n^>$. Singularities can appear only if two variables approach each
other, $z_i\rightarrow z_j$. We continue this function on the universal
covering
of $M_n^>$ which is
\a
M_n &=& \{(z_1,...,z_n): \quad z_i\neq z_j \quad if\quad i\neq j\}
\b
Choosing a point $P\in M_n^>$ and denoting by
$\gamma$ a path in $M_n$ starting at
$\gamma (0) =P$ and arriving at $\gamma (1) =z\in M_n$,
the expression $f_j (z\gamma )$
denotes the (unique) analytic continuation of $f_j$ from
$P\in M_n^>$ to $z\in M_n$ along $\gamma$.
We introduce an action ${\tilde\sigma}_i$ of the braid group
$B_n$ on functions $f_j$ by
\a
({\tilde\sigma}_if_j)(z) &=& f_j (z, \gamma_i (z))
\b
for a path $\gamma_i(z)$ in $M_n$ running from $P$ first interchanging
$P_i$ and $P_{i+1}$ in positive direction and then connecting
the resulting point to \\
$(z_1,...,z_{i-1},z_{i+1},z_i,z_{i+2},...,z_n)=
t_i(z)$ on a path in $t_iM_n^>$.\\
The interesting point is that the Artin relations (\ref{artin1}) can
be realized
on ${\tilde x}_i$ where
\a
({\tilde x}_i f_j) (z) &=& \int_{\gamma_i} f_j (t,z_1,...,z_{n-1})dt
\b
with $\gamma_i$ a loop around $z_i$.
Now the claim is that the relations
\a
{\tilde x}_j{\tilde\sigma}_i &=& \sum_k (B(i))_{jk} {\tilde x}_k
\label{commuting}
\b
for $ i=1,2,...,n$
are verified \cite{ConLue}, where $B(i)$ are
the braid matrices (\ref{represent}).
This allows us to get a representation of the braid groups on integrals
if the braid representation on the integrands is known.
The relation (\ref{commuting})
can be understood as commuting the generators of the braid group through
the integrals. Certainly this procedure can be iterated making possible
a computation of the braid and monodromy properties of generalized
hypergeometric integrals (in particular no ``charge condition"
is necessary). The graded structure which results from this construction
as well as details and specific computations are contained in
\cite{ConLue}.
Let us remark that in the two-dimensional conformal
perturbation theory braid properties of such generalized
hypergeometric integrals
without charge condition are needed. These problems were studied in
\cite{flu}.
Finally there are applications of the results and methods of this
paper in the theory of quantum groups. A step in this direction
was taken in \cite{LueTop,Lue} where the vertex construction
of Gomez and Sierra \cite{GomSie} was generalized
and put in the framework of an abstract braid module.
For some related ideas in the supergroup case $U_q gl(1,1)$ see \cite{kauff}.
\vskip.5cm
\vskip.3cm

{\bf Acknowledgements}
We would like to thank
G. Burde for having given some hints to us, M. L\"{u}dde, H.F. de Grote and R.
Flume
for useful discussions.

\end{document}